\documentclass[usenatbib]{mnras}

\usepackage{times,graphicx,amssymb,amsmath,natbib,color}

\newcommand{\cp}{\citep}
\newcommand{\ct}{\citet}

\title{Born Dry in the Photo-Evaporation Desert: {\it Kepler}'s Ultra-Short-Period Planets Formed Water-Poor.}
\author[Eric D. Lopez]{Eric D. Lopez,$^{1}$\thanks{E-mail: elopez@roe.ac.uk}
\\
$^{1}$SUPA, Institute for Astronomy, Royal Observatory Edinburgh, University of Edinburgh, Blackford Hill, Edinburgh, UK\\
}

\begin{document}

\date{Accepted 0000}

\label{firstpage}
\pagerange{\pageref{firstpage}--\pageref{lastpage}}
\maketitle

\begin{abstract}
Recent surveys have uncovered an exciting new population of ultra-short-period (USP) planets with orbital periods less than a day. These planets typically have radii $\lesssim$1.5 $R_{\mathrm{\oplus}}$, indicating that they likely have rocky compositions. This stands in contrast to the overall distribution of planets out to $\sim$100 days, which is dominated by low-density sub-Neptunes above 2 $R_{\mathrm{\oplus}}$, which must have gaseous envelopes to explain their size. However, on ultra-short-period orbits, planets are bombarded by intense levels of photo-ionizing radiation and consequently gaseous sub-Neptunes are extremely vulnerable to losing their envelopes to atmospheric photo-evaporation. Using models of planet evolution, I show that the rocky USP planets can easily be produced as the evaporated remnants of sub-Neptunes with H/He envelopes and that we can therefore understand the observed dearth of USP sub-Neptunes as a natural consequence of photo-evaporation. Critically however, planets on USP orbits could often retain their envelopes if they formed with very high-metallicity water dominated envelopes. Such water-rich planets would commonly be  $\gtrsim$2 $R_{\mathrm{\oplus}}$ today, which is inconsistent with the observed evaporation desert, indicating that most USP planets likely formed from water-poor material within the snow-line. Finally, I examine the special case of 55 Cancri e and its possible composition in the light of recent observations, and discuss the prospects for further characterizing this population with future observations.
\end{abstract}

\begin{keywords}
planets and satellites: atmospheres, planets and satellites: composition, planets and satellites: physical evolution
\end{keywords}


\section{Introduction}\label{introsec}

One of the greatest revelations from recent exoplanet surveys like NASA's {\it Kepler} mission has been the discovery of whole new classes of planets completely unlike anything found in the Solar System \cp{Borucki2010, Batalha2013}. In particular, we have uncovered an abundant new population of highly irradiated planets with sizes between that of Earth and Neptune and orbital periods $\lesssim$100 days \cp{Fressin2013,Petigura2013a}. This sample includes both a population of likely rocky super-Earths whose masses and radii indicate that they do not contain substantial gaseous envelopes \cp{Rogers2015} and are consistent with Earth-like compositions \cp{Dressing2015}, as well a large population of lower density sub-Neptunes whose sizes indicate that they must be covered with thick envelopes of volatile gases like water or hydrogen and helium (hereafter H/He). 

These super-Earth and sub-Neptune sized planets may be the most common type of planet in our galaxy \cp{Petigura2013b}, and they completely dominate the population of currently known exoplanets \cp{Batalha2013,Coughlin2016}, and yet there are no planets in our own Solar System in this size range or with such short orbital periods. As a result, understanding the nature and origin of these planets poses a key test to our models of planet formation, and will be essential to understanding how our Solar System fits into the broader context of low-mass planet formation. We know, for example, that the sub-Neptunes must have large gaseous envelopes that make up a large portion of their volume; however, we do not know, based on their masses and radii alone \cp{Rogers2010b}, the compositions of these envelopes. Are they solar composition H/He envelopes atop Earth-like rocky cores? If so then these envelopes would typically only need to make up a few percent of their total mass \cp{Lopez2014}, and it is possible that these planets could have formed close to their current irradiated orbits \cp{Lee2015b}. On the other hand, if these planets accreted most of their mass from water and other volatile ices, as is believed to be the case in Uranus and Neptune \cp{}, then they likely needed to migrate to their current orbits from beyond the snow-line \cp{Raymond2008}. This distinction is key to understanding the typical compositions of low-mass planets and the frequency of Earth-like planets with just the right amount of volatiles to be habitable \cp{Alibert2014}. 

A planet's composition, however, is not constant throughout its lifetime. Highly-irradiated planets, like most of those found by Kepler, are bombarded by large amounts of ionizing radiation in the extreme UV (hereafter simply referred to as XUV) \cp[e.g.,][]{Ribas2005,Sanz-Forcada2011}. At high levels of XUV irradiation this partially ionizes hydrogen high up in a planet's outer atmosphere, heating gas up to $\sim10^4$ K and creating a collisional Parker wind that can strip away mass from the planets \cp[e.g.,][]{Parker1958, Yelle2004, Murray-Clay2009}. This photo-evaporative wind has been directly observed in Lyman alpha for a handful of transiting hot Jupiters \cp[e.g.,][]{Vidal-Madjar2004,Lecavelier2004}, and was recently observed for the first time from a transiting hot Neptune GJ 436b \cp{Ehrenreich2015}. Likewise, photo-evaporation is believed to have stripped up to a terrestrial ocean of water from the atmosphere of early Venus \cp{Kasting1983,Watson1981}. Over billions of years, photo-evaporation can even completely strip the envelopes from Neptune sized planets, transforming them into bare rocky super-Earths \cp[e.g.,][]{Valencia2010, Lopez2012}. 

Moreover, there is significant observational evidence that photo-evaporation, or a comparable process, has sculpted the compositions and radii of the irradiated low-mass planet population. This is evidenced by a lack of lower density planets on more irradiated orbits which closely matches the predictions of photo-evaporation models from many different groups \cp[e.g.,][]{Lopez2014, Owen2013, Jin2014, Howe2015, Chen2016}. Indeed, since this "photo-evaporation threshold" for hot low-mass, low-density planet was first proposed in \ct{Lopez2012}, the number of low-mass planets with measured densities has more than tripled \cp[exoplanets.org,][]{Wright2011}, and yet this observed threshold remains valid \cp[e.g.,][]{Chen2016}. While it is possible that processes besides photo-evaporation, such as atmospheric erosion by impacts \cp[e.g.,][]{Catling2013, Inamdar2015, Schlicting2015, Liu2015, Inamdar2016}, have also played a significant role, what is clear is that the primordial compositions of short-period low-mass planets have been significantly modified by their post-formation evolution. 

In particular, if we look specifically at planets on Ultra-Short-Period (USP) orbits, i.e., those with periods $\lesssim$1 day, there is a dearth of sub-Neptune or Neptune sized planets \cp{Sanchis-Ojeda2014,Lundkvist2016}. \ct{Sanchis-Ojeda2014} first described this extreme population by examining 106 non-giant confirmed transiting planets and {\it Kepler} candidates with orbital periods of 4.2 to 24 hours. \ct{Sanchis-Ojeda2014} found an overall occurrence rate of 5.5$\pm$0.5 USP planets per thousand stars, with a sharp drop-off in planet occurrence at sizes larger than 1.4 $R_{\mathrm{\oplus}}$ and essentially no USP planets from 2-4 $R_{\mathrm{\oplus}}$. Subsequently, \ct{Lundkvist2016} confirmed a lack of planets in this size range on orbits receiving more than 650$\times$ the irradiation of the Earth for the subset of {\it Kepler} targets with stellar properties from asteroseismology and McDonald et al. (in prep) examined this planet occurrence desert as a function the lifetime integrated x-ray flux received by a planet and how this varies with stellar type.

This observed maximum size for USP planets roughly corresponds to the observed transition between rocky and non-rocky planets for planets with masses from radial velocity found by \ct{Rogers2015}, along with theoretical predictions for the maximum typical size of rocky super-Earths \cp[e.g.,][] {Mordasini2012b, Chiang2013}, indicating that this is a predominately bare rocky planet population without significant volatile envelopes. The one possible exception to this rule is 55 Cancri e, which I will discuss in more detail in Section \ref{55sec}. This absence of sub-Neptune sized USP planets is sometimes referred to as either the "sub-Neptune desert" or "super-Earth desert"; however, here I will refer to it as the "evaporation desert" under the assumption that it results from the complete stripping of gaseous envelopes from non-rocky sub-Neptune planets by atmospheric photo-evaporation, as should be expected from models of photo-evaporation for solar-composition H/He envelope \cp[e.g.,][]{Lopez2012,Owen2013,Chen2016}.

Critically, however, planets with different atmospheric compositions experience different levels of vulnerability to photo-evaporation. As a result, this means that we can use atmospheric evaporation as a tool to constrain the bulk compositions of the hot sub-Neptunes. In particular USP planets present a clean test since these planets generally cannot retain solar-composition envelopes. However, water vapor envelopes should be much more resistant to destruction by photo-evaporation. This is primarily because planets with water envelopes experience much less evolution in their radii as they cool and contract after formation \cp{Lopez2012}, their higher mean-molecular weight and oxygen abundance should produce additionally cooling in their photo-evaporative winds. Using a modified version of the planet evolution and evaporation model from \ct{Lopez2012}, \ct{Lopez2013}, and \ct{Lopez2014}, I will show that the observed photo-evaporation desert is readily reproduced if USP planets initially form as sub-Neptunes with solar composition H/He gaseous envelopes. If on the other hand, these planets had initially formed with large amounts of water then a large fraction of them should be able to retain their envelopes and have radii $\gtrsim$2 $R_{\mathrm{\oplus}}$, inconsistent the observed evaporation desert. Consequently, this suggests that the USP planets, at least, formed from water-poor material within their stars snow-lines.

\section{Modeling Planet Evolution and Photo-Evaporation}

This work uses a slightly modified version of the coupled thermal evolution and atmospheric photo-evaporation model previously described extensively in \ct{Lopez2012}, \ct{Lopez2013}, and \ct{Lopez2014}. Below is a brief summary of these models, along with a description of the modifications that have been made for modeling photo-evaporation from USP planets. The planets considered here will be low mass sub-Neptune and Neptune sized planets with massive volatile envelopes atop rocky cores. The cores are assumed to have an Earth-like mix of silicates and iron, as motivated by the observations of \ct{Dressing2015}, while the envelopes can be either solar composition H/He envelope or pure H$_2$O, i.e., "steam" envelopes. In reality, of course, a range of intermediate compositions are possible and very high metallicity envelopes will have species other than water, however, these two possibilities represent useful and commonly utilized end cases \cp[e.g.,][]{Rogers2010a, Nettelmann2011,Lopez2012} and water is by far the most important volatile ice for planets that form beyond the snow-line \cp{Marboeuf2008}.

Throughout this paper, I will be assuming that these planets orbit Sun-like stars, as is typically the case for the {\it Kepler} sample \cp{Batalha2013}, however, these results can be generalized to other stellar types as well by determining the XUV flux that planets receive as function of stellar type, which for FGK stars is strongly correlated with the incident bolometric used here \cp{Sanz-Forcada2011, Jackson2012}.

For the rocky cores of the planet, the model uses the ANEOS olivine \cp{Thompson1990} and SESAME 2140 Iron \cp{Lyon1992} equations of state, with an Earth-like 2:1 rock/iron ratio. For the gaseous envelopes the model uses the SCVH equation of state from \ct{Saumon1995} for solar-composition H/He envelopes and H2O-REOS \cp{Nettelmann2008, French2009} for water. Just to be clear, the gaseous envelopes considered here are not thin atmospheres like that on Earth. These are extremely thick convective envelopes like those found on the ice and gas giants of the solar system. These envelopes should be fully convective adiabats reaching pressures of up to $\sim$ 1 Mbar, and often contain the bulk of the planet's volume. Finally atop the gaseous envelope is a relatively small radiative atmosphere, which is assumed to be isothermal at the planet's equilibrium temperature.

In order predict planetary radii, masses, and compositions and how these quantities vary over time, it is necessary to use a model of planetary structure and evolution. There are three main components to this model. First the structure model uses the equations of state described above to solve for the hydrostatic structure of a planet's core and gaseous envelope as a function planetary mass, core/envelope fraction, external irradiation, and the entropy of its convective adiabat in the gaseous envelope. Next the thermal evolution model then evolves the structure models in time, beginning after the planet finishes accreting its envelope, and tracks how a planet cools and contracts as it radiates away its initial heat from formation. Finally, the atmospheric photo-evaporation model tracks how a planet's envelope mass, and therefore radius, changes as it loses mass due to hydrodynamic atmospheric evaporation driven by XUV radiation. 

Both the thermal evolution and the photo-evaporative evolution are calculated simultaneously so that a planet's composition and radius are solved for self consistently as it evolves in time. As a general rule, however, while the thermal evolution has a large impact on the evaporative mass loss rate and mass loss certainly impacts planetary radii, atmospheric mass loss generally does not have a large impact on the thermal state, i.e., entropy, of the planetary interior \cp{Lopez2013,Chen2016}. The only time this is not true is when mass loss is so vigorous that the mass-loss timescale becomes short compared to the thermal timescale, in which case planets generally lose their envelopes completely \cp{Lopez2012}.

\subsection{Thermal Evolution}

The thermal evolution model is summarized in Equation (\ref{thermaleq}).

\begin{equation}\label{thermaleq}
\int_{M_{\mathrm{core}}}^{M_{\mathrm{p}}} dm \frac{T dS}{dt} = - L_{\mathrm{int}} + L_{\mathrm{radio}} - c_{\mathrm{v}} M_{\mathrm{core}}  \frac{dT_{\mathrm{core}}}{dt}
\end{equation}

The left-hand side describes the cooling and contraction of a planet's convective envelope, where $M_{\mathrm{core}}$ is the mass of the planet's rocky core, $M_{\mathrm{p}}$ is the total planet mass, S is the entropy of the convective adiabat in the volatile envelope. The right-hand side meanwhile describes the various energy sources and sinks that control a planets thermal evolution. $L_{\mathrm{int}}$ is the intrinsic luminosity from radiative cooling from the atmosphere, calculated using the results of the radiative transfer models of \ct{Fortney2007} assuming a solar metallicity for H/He envelopes, and 50$\times$ solar for steam envelopes. $L_{\mathrm{radio}}$ meanwhile, describes heating by radioactive decay in the planet's rocky core, assuming Earth-like abundances of Potassium, Thorium, and Uranium. Finally, $c_{\mathrm{v}}=0.75 \, J \, K^{-1} \, g^{-1}$ \cp{Lopez2014} is the heat capacity of the rocky core, and this term accounts for the cooling of the core in conjunction with the envelope. Each of these terms is discussed in great detail in \ct{Lopez2014}. 

\subsection{Photo-evaporation Model}\label{evapsec}

When modeling evaporation, there are three key points in a planet's upper atmosphere that are key to determining the type of mass loss that planets will experience. First is the XUV photosphere, i.e., the point where the planet becomes optically thin to ionizing photons. Here this is the base of the photo-evaporative wind. Second, is the sonic point, if one exists, where the evaporative wind becomes transsonic and therefore unbound to the planets. Lastly, there is the exobase, which is the point at which the mean free path becomes comparable to the scale height and the evaporative wind ceases to be collisional. For highly irradiated planets such as those considered here, the XUV photosphere and sonic point both occur well before the exobase and so the evaporative wind is fully collisional up to the point where it becomes becomes unbound from the planet, and so the wind is said to be fully collisional or hydrodynamic, and it is governed by the equations of fluid dynamics \cp{Murray-Clay2009,Owen2012}. 

In between the XUV photosphere and the sonic point, the evaporative wind is able to cool and radiate away much of the heating from ionization and it is this cooling which sets the overall efficiency of mass loss, i.e., the fraction of the XUV heating that goes into performing useful work to remove mass from the planet's atmosphere. To determine these mass loss rates, here I calculate mass loss rates for two relevant limits, energy-limited and recombination-limited escape, as discussed below in Sections \ref{elsec} and \ref{radsec}, following the formalism of \ct{Murray-Clay2009} as subsequently modified by \ct{Chen2016}. I then use whichever of these two rates is smaller as in \ct{Jin2014} and \ct{Chen2016}. Here I have made an additional modification to the equations from \ct{Chen2016}, by making the dependence on mean molecular weight both in and below the photo-evaporative wind explicit, since I will be considering both H/He and steam envelopes. 

\subsubsection{Energy Limited Escape}\label{elsec}

One particularly useful, and widely utilized, limit appropriate for most hot neptunes with orbital periods around $\sim$10 days is energy-limited escape \cp[e.g.,][]{Watson1981, Baraffe2006, Murray-Clay2009, Jackson2010, Valencia2010, Lopez2012, Jin2014, Chen2016}. In this regime, however, the x-ray photosphere is at a temperature such that the dominant coolants are atomic line coolants, particularly carbon and oxygen \cp{Owen2013}, and so the efficiency of photo-evaporative mass loss is more or less constant as a function of the incident flux \cp{Murray-Clay2009}. To clarify, these cooling lines carry away most of the energy in the evaporative wind, and only a fraction goes into useful work removing mass from the atmosphere. The escape rate in this limit is described as "energy-limited", however, because the mass-loss rate is linear with the incident XUV flux, and so the energy that goes into mass loss is roughly proportional to the energy received at the planets XUV photosphere. The energy-limited mass loss rate is described by Equation (\ref{eleq}), taken from \ct{Erkaev2007}.

\begin{equation}\label{eleq}
\dot{M}_{\mathrm{EL}}  = -\frac{\epsilon_{\mathrm{XUV}}  \pi F_{\mathrm{XUV}} R_{\mathrm{base}}^3 }{G M_{\mathrm{p}} K_{\mathrm{tide}} }
\end{equation}

Here $\epsilon_{\mathrm{XUV}}$ is the parameterization of the efficiency of photo-evaporation, that is the fraction of the incident XUV heating that goes into removing mass, rather than being radiated away. Based on the results of XUV radiative transfer calculations \cp[e.g.,][]{Murray-Clay2009,Owen2012} this is usually taken to be $\sim$10\% for solar composition atmospheres \cp[e.g.,][]{Jackson2010, Valencia2010, Lopez2012, Jin2014, Chen2016}. Moreover, evaporation models using an efficiency of $\sim$10\% are able to reproduce the observed photo-evaporation threshold \cp{Lopez2012,Lopez2014}. In reality of course, this assumption of constant efficiency is only a rough approximation, the photo-evaporation efficiency can easily vary from ~5-20\%, particularly as a planet's gravity varies as it loses mass and contracts \cp{Owen2013}. Nonetheless, $\epsilon_{\mathrm{XUV}}\sim$10\% is a useful approximation and in any case, the results presented here are sufficiently robust that they are insensitive to any modest variation in $\epsilon_{\mathrm{XUV}}$. 

It is worth pointing out, however, that the efficiency of evaporation should be significantly lower for higher metallicity atmospheres, first because the higher mean-molecular weight will reduce the scale height in the evaporative wind,  thereby increasing the number of scale heights between the XUV photosphere and the sonic point, and second because in this regime metal atomic lines are the dominant coolants. Both these effects should lead to additional cooling in the photo-evaporative wind and will reduce the evaporation efficiency. Indeed \ct{Ercolano2010}, examined this affect for photo-evaporation in proto-planetary disks and found that $\epsilon \propto Z^{-0.77}$. Applied to pure water atmospheres, this would suggest an efficiency closer to $\epsilon_{\mathrm{XUV}}\sim$1\%. In Section \ref{resultssec} I examine the effect of such a reduction in the mass loss rates for water atmospheres, however, I find that the key results presented here are generally insensitive to such a change. 

As for the rest of Equation (\ref{eleq}), $F_{\mathrm{XUV}}$ is the XUV flux at the planet's age and orbit, calculated using the results from \ct{Ribas2005} for Sun-like stars. $K_{\mathrm{tide}}$ is a correction factor that accounts for the fact that material only needs to escape to the planet's tidal roche lobe, not to infinity. Finally, $R_{\mathrm{base}}$  is the planet radius at the XUV photosphere, i.e., at the base of the evaporative wind, calculated using Equation \ref{rbaseeq}.

\begin{equation}\label{rbaseeq}
R_{\mathrm{base}} \approx R_{\mathrm{p}} + H_{\mathrm{below}} \ln \left( \frac{P_{\mathrm{photo}} } {P_{\mathrm{base}} } \right)
\end{equation}

Here $H_{\mathrm{below}} = \left( k_{\mathrm{B}} T_{\mathrm{eq}} \right) / (\mu_{\mathrm{below}} m_{\mathrm{H}} g) $ is the scale height in the regime between the optical and XUV photospheres, which is assumed to be roughly isothermal at the equilibrium temperature $T_{\mathrm{eq}}$ and $\mu_{\mathrm{below}}$ is the mean molecular weight in this region, assumed to by 2.5 for H/He envelopes, and 18 for steam envelopes. $P_{\mathrm{photo}} \approx $20 mbar is the pressure level for the optical photosphere for a transit geometry \cp{Fortney2007}, while $P_{\mathrm{base}} \approx (m_{\mathrm{H}} g) / (\sigma_{ \mathrm{ \nu 0}})$, is the pressure level for XUV photosphere, typically  $\sim$1 nbar.  This is calculated using $\sigma_{\mathrm{\nu 0}} = 6 \times 10^{-18} (h \nu_{\mathrm{0}} / \mathrm{13.6 eV} )^{-3}$ cm$^2$, with $h /nu_{ \mathrm{0}}$ = 20 eV as the typical energy for the ionizing photons \cp{Murray-Clay2009}.

\subsection{Radiation-Recombination Limited Escape}\label{radsec}

At Ultra-Short-Period orbits, however, there is another limit that becomes extremely important. On such highly irradiated orbits, the temperature of photo-evaporative wind and the ionization fraction become high enough, that Hydrogen recombination and the subsequent radiation of Lyman series photons becomes the dominant coolant \cp{Murray-Clay2009}. In this case the evaporative wind can be approximated by the classical Parker wind solution \cp{Parker1958}, with an isothermal wind at $\sim10^4$ K. In that case, the evaporative mass-loss rate will be given by Equation (\ref{rreq1}), taken from \ct{Murray-Clay2009}.

\begin{equation}\label{rreq1}
\dot{M}_{\mathrm{RR}}  = -4 \pi \rho_{\mathrm{s}} c_{\mathrm{s}} R_{\mathrm{s}}^2
\end{equation}

Where, $R_{\mathrm{S}}$ is the radius of the sonic point, $\rho_{\mathrm{s}}$ is the density at the sonic point, and $c_{\mathrm{s}} = (k_{\mathrm{B}} T_{\mathrm{wind}} /  (\mu_{\mathrm{wind}} m_{\mathrm{H}}))^{1/2}$ is the sound speed at the sonic point.  Here the radius of the sonic point is given by $R_{\mathrm{S}} = G M_{\mathrm{p}} / (2 c_{\mathrm{s}}^2)$, unless this gives a value that is smaller than $R_{\mathrm{base}}$, in which case I set $R_{\mathrm{S}}=R_{\mathrm{base}}$. The sound speed  meanwhile is set by assuming $T_{\mathrm{wind}}\approx10^4$ K and that $\mu_{\mathrm{wind}}=0.62$ for H/He envelopes and 3 for steam envelopes, since the hydrogen should be mostly ionized, and for steam atmospheres the oxygen should be singly ionized. Finally, the density at the sonic point is given by Equation (\ref{rhoeq}). 

\begin{equation}\label{rhoeq}
\rho_{\mathrm{s}}  \approx \rho_{\mathrm{base}}  \exp{\left[ \frac{ G M_{\mathrm{p}} }{ R_{\mathrm{base}} c_{\mathrm{s}}^2 } \left( \frac{ R_{\mathrm{base}} }{ R_{\mathrm{s}} } - 1 \right)  \right]}
\end{equation}

Where $\rho_{\mathrm{base}} =  n_{\mathrm{+,base}} \mu_{\mathrm{+,wind}} m_{\mathrm{H}}$ is the density at the base of the wind, and $n_{\mathrm{+,base}}$ is the density of ions at the base of the wind, which is found by setting $\alpha_{\mathrm{rec,B}} n_{\mathrm{+,base}}^2 \approx  (F_{\mathrm{XUV}} / h \nu_{ \mathrm{0}} ) \sigma_{\mathrm{\nu 0}} n_{\mathrm{0,base}} $. Here $\mu_{\mathrm{+,wind}}$ is the mean molecular weight of ions in the wind, 1.3 or 6 for H/He or steam atmospheres; $\alpha_{\mathrm{rec,B}}$ is the case B recombination rate, and $n_{\mathrm{0,base}} \approx 1 / (\sigma_{\mathrm{\nu 0}} H_{\mathrm{base}} ) = G M_{\mathrm{p}} / (\sigma_{\mathrm{\nu 0}} c_{\mathrm{s}}^2 R_{\mathrm{base}}^2)$ is the neutral density at the base of the wind.

Putting all of this together gives Equation (\ref{rrtotaleq}), which is similar to Equation (3) in \ct{Chen2016}, except that here I have made explicit the dependence on the mean molecular weight, and therefore composition of the atmosphere.

\begin{equation}\label{rrtotaleq}
\begin{aligned}
\dot{M}_{\mathrm{RR}}  = {} &-4 \pi c_{\mathrm{s}} R_{\mathrm{s}}^2  \mu_{\mathrm{+,wind}} m_{\mathrm{H}} \left( \frac{F_{\mathrm{XUV}} G M_{\mathrm{p}}  }{h \nu_{ \mathrm{0}} \alpha_{\mathrm{rec,B}}  c_{\mathrm{s}}^2 R_{\mathrm{base}}^2 }  \right)^{1/2} \\
& \times \, \exp{\left[ \frac{ G M_{\mathrm{p}} }{ R_{\mathrm{base}} c_{\mathrm{s}}^2 } \left( \frac{ R_{\mathrm{base}} }{ R_{\mathrm{s}} } - 1 \right)  \right]}
\end{aligned}
\end{equation}

\section{Results for USP Planets}\label{resultssec}

With this model in place, I can now explore the possible evaporation histories of USP planets and how their final compositions and radii depend on their mass, irradiation, and initial compositions. To do this I ran a large suite of model evolution runs designed to explore this parameter space. This suite of models is similar to those described in \ct{Lopez2013}, with three key differences. First, I have made the modifications to the photo-evaporation prescription described above in Section \ref{evapsec}. Second, I have extended these models up to higher levels of irradiation $>$1000 $F_{\mathrm{\oplus}}$ relevant for for USP planets. Finally, I have also run the models for pure steam envelopes in order to compare these to the results for solar composition envelopes.

To explore this parameter space, I ran over 20,000 individual models covering a wide range of planet core masses, orbits, and initial envelope fractions for each envelope type and mass-loss prescription. These models were run on a grid with core masses spaced uniformly in log space from 1 to 20 Earth masses and present-day bolometric incident flux spaced log uniform from 100 to 10,000 $\times$ the insolation of the Earth, for comparison a planet on a 1 day orbit around a Sun-like star receives $\approx$ 2600 $F_{\mathrm{\oplus}}$. For planet's with H/He envelopes, I spaced the initial envelope mass fractions uniformly in log space from 0.1 to 50\% of the total initial mass. Note that only planets with final envelope fractions $\lesssim10\%$ will enter the radius regime that we care about here below 4 $R_{\mathrm{\oplus}}$, therefore including planets with larger initial envelope fractions will not significantly affect the results presented here. Such planets will typically either retain most of their initial envelopes and end up with sizes $\gtrsim$6 $R_{\mathrm{\oplus}}$ or else lose their envelopes completely and add slightly to the population of stripped rocky planets. 

I have chosen to space the grid points uniformly in log space partly because this allows us to uniformly explore the relevant parameter space, and partly because such a spacing, at least for the mass and initial composition distributions, is somewhat motivated by observations. For example, \ct{Mayor2011} examined the distribution of planet masses found by radial velocities with orbital periods $<$100 days and found that the key break in the mass distribution occurs at $\sim15-30$ $M_{\mathrm{\oplus}}$. Above this value, the planet occurrence rate is comparatively low, while below $\sim15-30$ $M_{\mathrm{\oplus}}$  the occurrence rate is quite high down to at least $\sim1$ $M_{\mathrm{\oplus}}$. Admittedly, this mass distribution is still very uncertain, and it is possible that there is a modest enhancement in the occurrence rate around $\sim2$ $M_{\mathrm{\oplus}}$, but to first order log-uniform in core mass up to $\sim20$ $M_{\mathrm{\oplus}}$ seems like a reasonable approximation. 

Likewise, a log-uniform distribution of initial envelope fractions is also a reasonable starting point. This is roughly what is found by population synthesis models for the initial accretion of gaseous envelopes for non-giant planets \cp{Mordasini2012b}. Moreover, using a evolution and photo-evaporation prescription similar to the one used here \ct{Chen2016} recently found that an initial distribution that is log-uniform in planet mass and initial H/He envelope fraction does an excellent job of reproducing the current radius distribution of short-period planets when evaporation is taken into account. Moreover, the results presented here are sufficiently robust that they should be insensitive to any reasonable variation in the initial distribution used.

\begin{figure}
  \begin{center}
    \includegraphics[width=3.5in,height=2.5in]{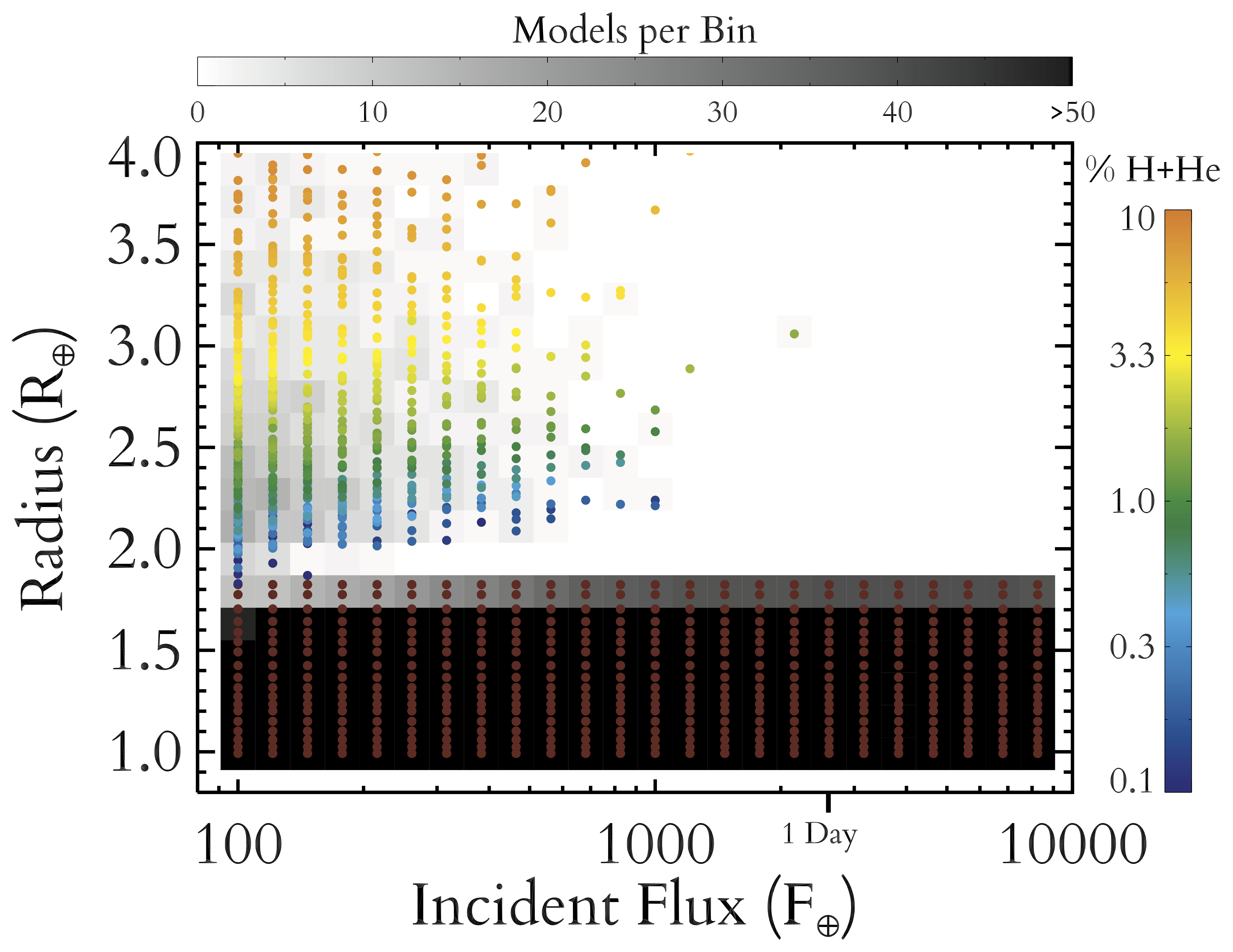}
  \end{center}
  \caption{The final planet radius predicted by the model after 5 Gyr of thermal and photo-evaporative evolution vs. the incident bolometric flux that a planet receives at it's orbit, for planets with solar composition H/He envelopes atop Earth-like cores. $\gtrsim$20,000 individual model runs where performed to generate this figure. The results of individual runs are shown by the points, which have been color-coded by their final H/He envelope mass fraction. Rust-colored points in the bottom right indicate bare rocky planets which have completely lost their H/He envelopes. The grey-scale background meanwhile shows the number of models that ended up in each radius-flux bin, where darker shades corresponds to a higher density of points, and clear regions correspond to areas devoid of models. On the x-axis, I have also indicated the incident flux received by a planet on a 1 day orbit around a 5 Gyr old Sun-like star. \label{hhefig}}
\end{figure}

\begin{figure}
  \begin{center}
    \includegraphics[width=3.5in,height=2.5in]{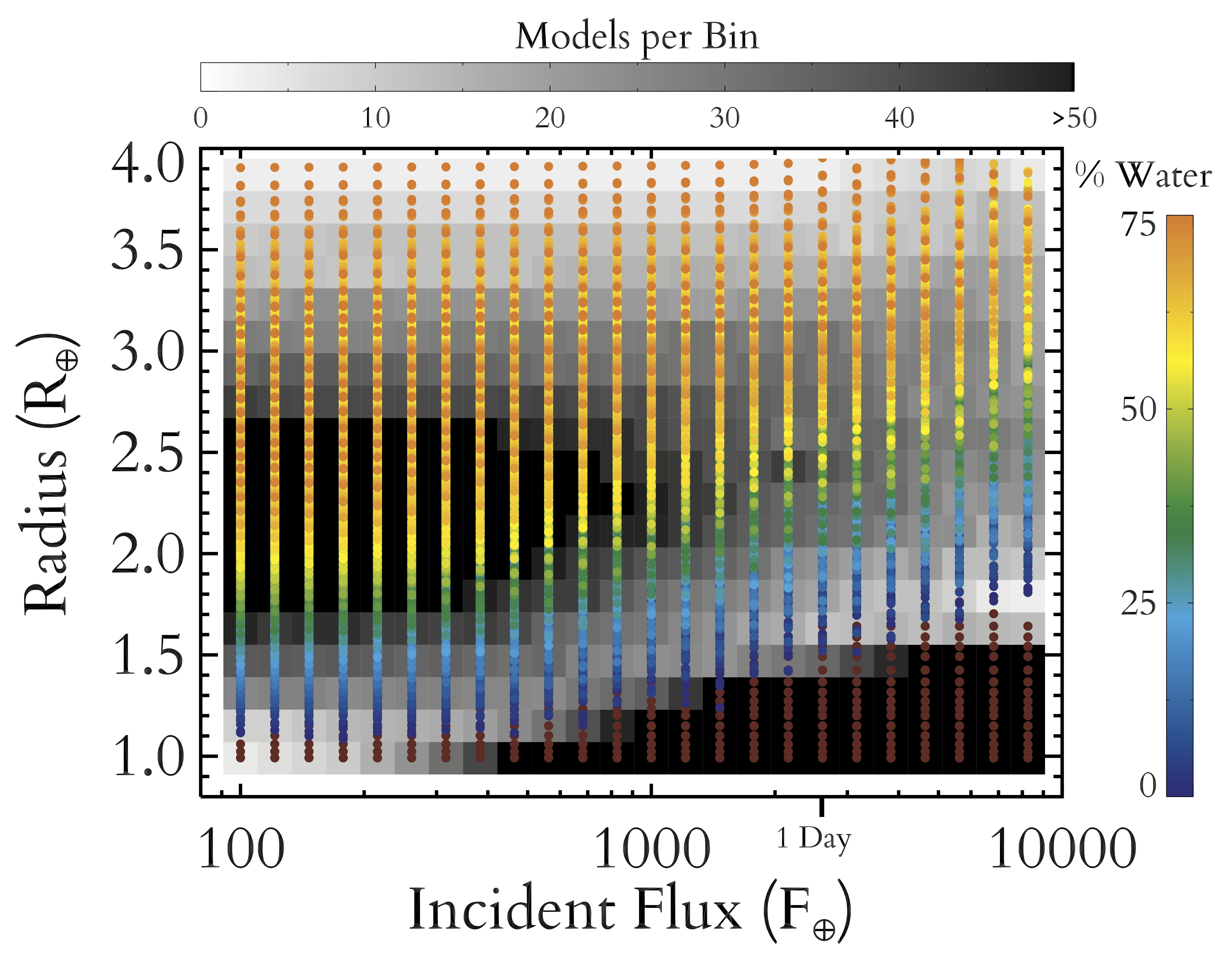}
  \end{center}
  \caption{Similar to Figure \ref{hhefig}, except here I show the results for models with pure steam envelopes. Unlike for H/He envelopes, here I find a large number of models that retain substantial water envelopes, despite being on USP orbits.  \label{waterfullfig}}
\end{figure}

Figure \ref{hhefig} shows the results of this suite of models for planets with solar composition H/He envelopes atop Earth-like rocky cores. Note that the final planetary radii are primarily determined the final H/He envelope fractions \cp{Lopez2014}, which are in turn strongly sculpted by photo-evaporation according to the planet's initial masses, orbits, and compositions \cp{Lopez2013}. Beyond this, however, there is relatively little dependence on initial conditions since the photo-evaporation history of a planet only affects the thermal state if it's envelope in cases of extreme evaporation, which generally end in a planet completely losing its envelope \cp{Lopez2013,Owen2016a,Chen2016}. Here we can clearly see that it is extremely difficult for any non-giant planets to retain such envelopes on orbits that receive  $\gtrsim1000$ $F_{\mathrm{\oplus}}$, and the models predict that no low-mass planets should retain H/He envelopes on USP orbits, consistent with the observed desert seen by \ct{Sanchis-Ojeda2014} and \ct{Lundkvist2016}. 

\begin{figure}
  \begin{center}
    \includegraphics[width=3.5in,height=2.5in]{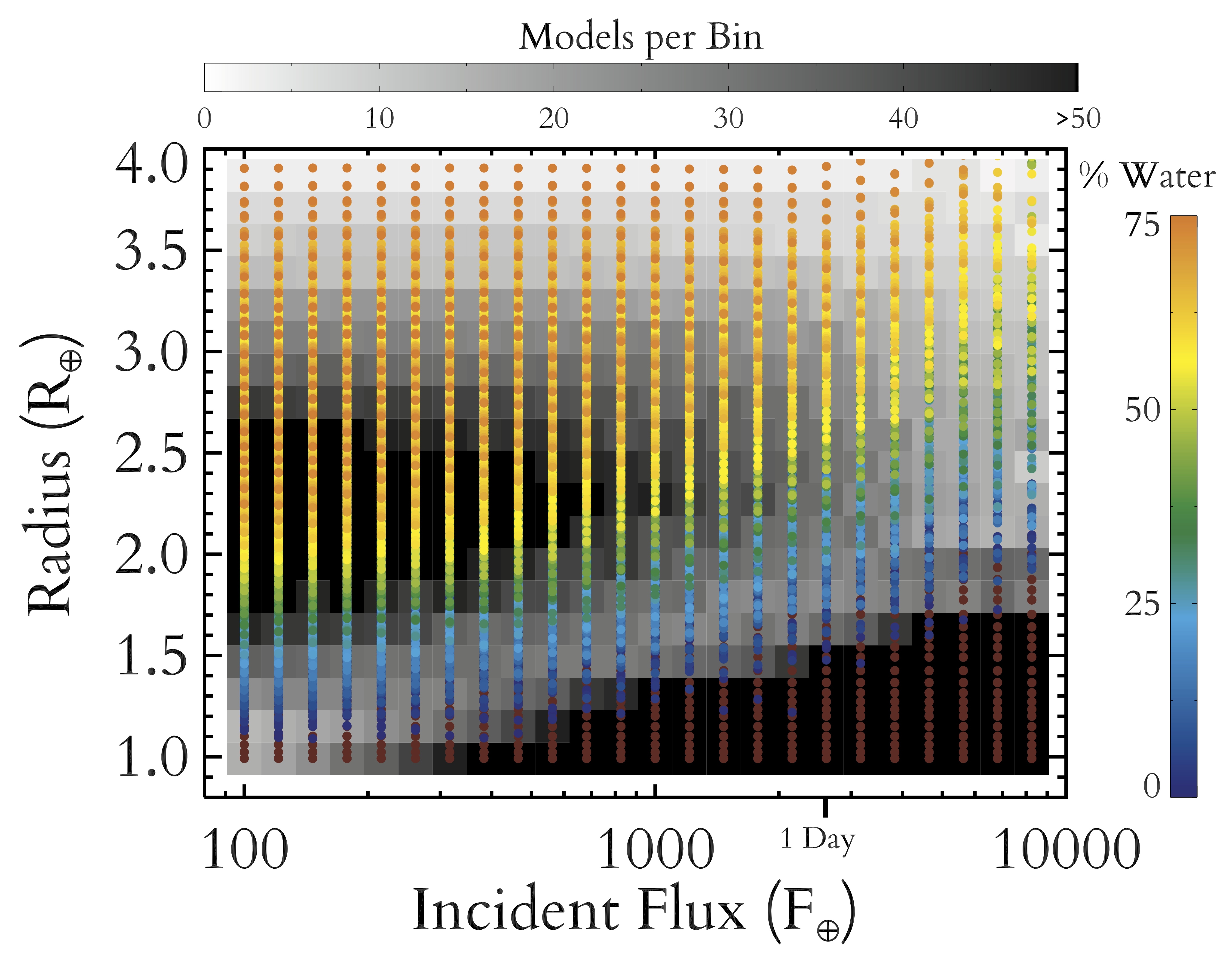}
  \end{center}
  \caption{Similar to Figure \ref{waterfullfig}, except that here I have used the same mass loss prescription as in Figure \ref{hhefig}. This provides an upper bound on the likely mass loss rate for pure steam atmospheres and shows that the resilience of water envelopes to photo-evaporation is primarily due to differences in their composition and thermal evolution.  \label{waterhhefig}}
\end{figure}

However, the results are very different when we turn to planets with large water envelopes, shown in Figures \ref{waterfullfig} and \ref{waterhhefig}. The setup for these models is very similar to that used for the H/He envelopes in Figure \ref{hhefig} and I have used the same log-uniform grid of core masses and irradiations. However, for water envelopes, a log-uniform distribution of initial envelope fractions is no longer appropriate. Rather than accreting their envelopes in the gas phase directly from the proto-planetary disk, water envelopes would instead form from collisions with water rich icy planetesimals from beyond the snow-line \cp[e.g.,][]{Raymond2008,Alibert2013}. Based on the results of population synthesis models \cp[e.g.,][]{Mordasini2012b,Alibert2013}, for the initial water envelope fraction distribution I use a uniform distribution from 0-75\% water by mass, where this maximum is based on the maximum possible water mass fraction produced by planet collisions  according to \ct{Marcus2010b}.

Figure \ref{waterfullfig} shows the results for the grid of models using the photo-evaporation model described in Section \ref{evapsec} with appropriate mean-molecular weights for water envelopes and $\epsilon_{\mathrm{XUV}}=1\%$. Here, unlike in Figure \ref{hhefig}, there are a large number of models which can hold on to significant water envelopes even on orbits that receive $\sim$10000 $F_{\mathrm{\oplus}}$, corresponding to orbital periods of $\sim$9 hours for Sun-like stars. This is clearly inconsistent with the observed evaporation desert \cp{Sanchis-Ojeda2014,Lundkvist2016}, and seems to rule out the possibility that such planets have contributed significantly to USP population found by {\it Kepler}.

Moreover, this result should be insensitive to reasonable variations in the photo-evaporation model used here. To test the robustness of this result, I re-ran the water envelope evolution grid, but used the same evaporation model used to create Figure \ref{hhefig}, i.e., I set the mean-molecular weight back to the values for H/He envelopes and set $\epsilon_{\mathrm{XUV}}=10\%$, but applied this to water envelopes.  The result is shown in Figure \ref{waterhhefig} and although this does somewhat enhance the amount of evaporation that takes place, the main results are unchanged. 

There are two main reasons for the resilience of water envelopes against destruction by photo-evaporation. First, compared to H/He envelopes, water envelopes experience much less evolution in their radii due to thermal evolution \cp{Lopez2012}. For example, a planet with a $\sim1\%$ H/He envelope atop an Earth-like core and a planet with a $\sim50\%$ water envelope atop that same core will each be $\sim 2$ $R_{\mathrm{\oplus}}$ at a few Gyr \cp{Lopez2012,Lopez2014}. However, even in the absence of any evaporation, the planet with the H/He envelope would have been $\gtrsim 6$ $R_{\mathrm{\oplus}}$ at 10 Myr, while the planet with the water envelope will only be $\sim 3$ $R_{\mathrm{\oplus}}$ \cp{Lopez2012}. By itself, this large difference in radii when the planets were young, and the XUV irradiation was high, means that H/He envelopes are almost an order of magnitude more vulnerable to photo-evaporation than water envelopes \cp{Lopez2012}. Moreover, as this example shows, to match the planet's radius a sub-Neptune-sized planet with a pure water envelope would need to have much more of the planet's mass in the water envelope than in a H/He envelope. As a result, to completely shed their volatile envelopes, planets with water envelopes need to lose more than an order of magnitude more mass while being about an order of magnitude less vulnerable to evaporation. As a result, water envelopes should be much more resistant to destruction by photo-evaporation than H/He envelopes, and unless one assumes an extremely high evaporation efficiency, than they should often be able to survive on USP orbits.

\subsection{The Curious Case of 55 Cancri e}\label{55sec}

\begin{figure}
  \begin{center}
    \includegraphics[width=3.5in,height=2.5in]{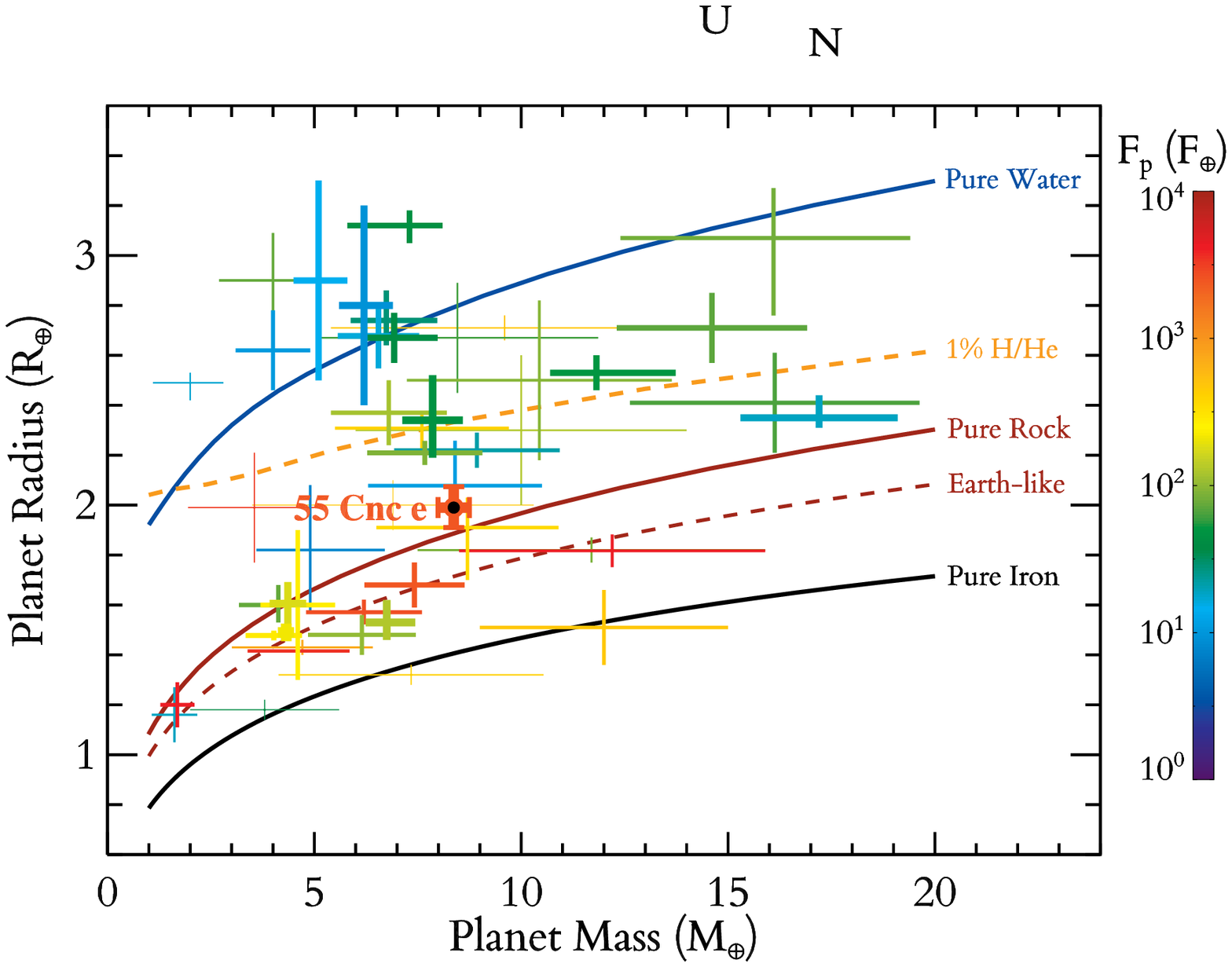}
  \end{center}
  \caption{The planetary mass-radius diagram for all currently known exoplanets with masses measured to at least 50\% precision. Points are color-coded according to the irradiation they receive relative to the incident flux the Earth receives from the Sun. To highlight the points with the most precise mass measurements I have scaled the thickness of the error bars inversely with the log of the precision of the mass measurement. In addition, I have plotted predicted mass-radius curves using the models of \ct{Lopez2014}. Going up from the bottom, the black solid curve is for pure iron, the dashed rust-colored curve is for a Earth-like mixture of rock and iron, the solid rust-colored curve is for pure silicate planets, the dashed orange curve is for a 1\% solar-composition H/He envelope atop an Earth-like core, and the solid blue curve is for a pure water planet.  Both the 1\% H/He and pure water curves assume an age of 5 Gyr and an irradiation of 100 $F_{\mathrm{\oplus}}$. 55 Cancri e, is the highlighted point sitting just above the pure silicate curve. \label{mrfig}}
\end{figure}

As mentioned in Section \ref{introsec}, there is one possible and well known exception to the general observation that all non-giant USP planets are consistent with bare rocky compositions and that is 55 Cancri e. Initially discovered by radial velocity with a mass of $8.3\pm0.3 \, M_{\mathrm{\oplus}}$ and a period of 0.736 days \cp{Fischer2008,Dawson2010}, 55 Cancri e was subsequently discovered to be transiting with both the Spitzer and MOST space telescopes \cp{Demory2011,Winn2011}, with a radius of 2.00$\pm0.14 \, R_{\mathrm{\oplus}}$ in the visible \cp{Winn2011} and $2.08\pm^{0.16}_{0.17} \, R_{\mathrm{\oplus}}$ in the infrared \cp{Demory2011}. A subsequent, joint analysis of both data sets along with additional Spitzer observations by \ct{Demory2012} gave a radius of $2.17\pm0.14 \, R_{\mathrm{\oplus}}$, which appeared to firmly rule out a bare rocky composition for the planet, indicating that it must have a significant volatile envelope. Since the photo-evaporation timescale for a H/He envelope on this planet would be on the order of a few Myr \cp{Valencia2010}, it was inferred that the planet must have a substantial steam envelope comprising $\sim20-30\%$ of the planet's mass. Using a bayesian comparison to planetary structure models \cp{Dorn2016a}, similar to those used here, \ct{Dorn2016b}, showed that 55 Cancri e is inconsistent with an Earth-like composition unless it somehow has a smaller radius $\approx$1.75 $R_{\mathrm{\oplus}}$

However, subsequent studies have since revised the radius of the planet downward, first to $1.99\pm^{0.084}_{0.080} \, R_{\mathrm{\oplus}}$ \cp{dragomir2014} and most recently down to $1.92\pm0.08 \, R_{\mathrm{\oplus}}$ \cp{Demory2016a}. Figure \ref{mrfig} plots the current mass and radius of 55 Cancri e from \ct{Demory2016a}, along with other planets in this mass range against theoretical mass-radius relations from \ct{Lopez2014}. While this revised radius does place the planet within 1$\sigma$ of the mass-radius curves for silicate rocky planets without iron cores, it is still $\approx2.0\sigma$ inconsistent with an Earth-like composition of 2/3 silicate mantle and 1/3 iron core, even using the updated rocky mass-radius relation from \ct{Zeng2016}. Using these models, the current mass and radius would imply a core mass fraction $\lesssim$15\%. However, there are no known mechanisms for producing planets this massive without iron cores, indeed planetary collisions if anything should increase the iron mass fraction \cp{Marcus2010a}. Moreover, we would probably expect an Earth-like core-mantle composition based on the stellar abundances of iron, silicon, and magnesium, which \ct{Brewer2016} found were all similarly enhanced relative to the Sun by 0.3-0.4 dex. However, the stellar abundances of 55 Cancri may warrant further attention from observers since when \ct{Dorn2016b} reviewed the range of compositions allowed by previous studies they found that [Mg/Fe] could range from -0.33 to 0.51 and [Si/Fe] could range from -0.38 to 0.28, when accounting for iron-rich and iron-poor cases.

\begin{figure}
  \begin{center}
    \includegraphics[width=3.5in,height=2.5in]{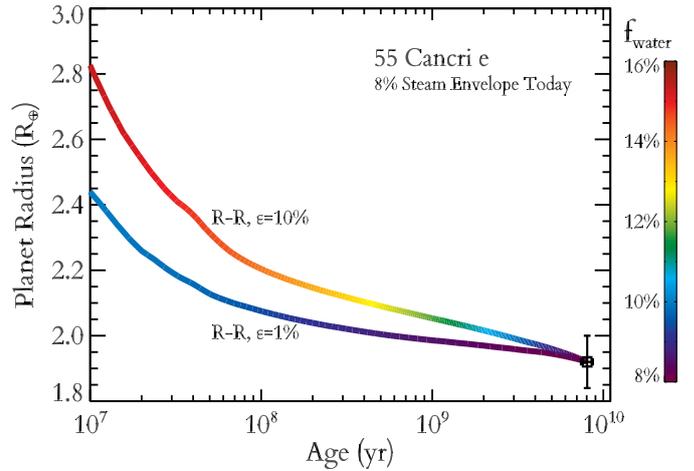}
  \end{center}
  \caption{Possible evolution histories for photo-evaporating steam envelopes on 55 Cancri e. Here I have plotted planet radius as a function of age for two different possible evaporation models, with the curves color-coded by the fraction of their total mass in the water envelope at that age. The black point at the bottom right shows the current planet radius and stellar age for 55 Cancri e from \ct{Demory2016a}. The top curve corresponds to an optimistic case for photo-evaporation assuming that the mass loss rate is given either by the radiation-recombination limited rate from Equation (\ref{rrtotaleq}) or the energy-limited rate from Equation (\ref{eleq}) with $\epsilon_{\mathrm{XUV}}=10\%$. The bottom curve, meanwhile, uses a more realistic value of $\epsilon_{\mathrm{XUV}}=1\%$. In both cases, the models are able to match the current estimate for the water envelope fraction of $\approx$8\% with a reasonable initial envelope fraction of 10-15\%. \label{cancrifig}}
\end{figure}

If, for now, we assume that 55 Cancri e must have an Earth-like core and that the current mass and radius are correct, than I find that it must have an water envelope comprising $8\pm3\%$ of the total planet mass to match the mass and radius. This may not sound like much, but it would correspond to a pressure $\sim660$ kbar at the bottom of the water envelope, which for comparison is $\approx600\times$ the pressure at the bottom of the Marianas trench and is more than enough to completely hide any rocky surface. 

Furthermore, given it's relatively large core mass a water envelope on 55 Cancri e should be fairly resistant to destruction by atmospheric photo-evaporation. Using the water envelope mass-loss model presented in Figure \ref{waterfullfig}, with both realistic and optimistic values of $\epsilon_{\mathrm{XUV}}$ in the energy-limited regime, I examined the possible evaporation history of 55 Cancri e. The predicted evolution histories for the planet radius and water envelope mass fraction are shown in Figure \ref{cancrifig}. Here I find that 55 Cancri e can retain the current predicted 8\% envelope if it had an initial water mass fraction of $\approx10-15\%$, depending on the value of $\epsilon_{\mathrm{XUV}}$ used, and that the current envelope should be resistant to destruction by further evaporation going forward. These are perfectly reasonable initial conditions, if 55 Cancri e formed from water rich material from beyond the snow-line \cp{Raymond2008}.

 This scenario may also be consistent with the observations of \ct{Demory2016b}, which measured the day night temperature difference on 55 Cancri e, and found that while the nightside is about half the temperature of the dayside, $1380\pm400$ K vs. $2700\pm270$ K, it is also still quite hot. The high temperature of the nightside may indicate the presence of an atmosphere that is recirculating energy from the dayside, although models suggest that this could also be achieved by tidal heating it the planet has a modest eccentricity \cp{Barnes2010,Bolmont2013}. Meanwhile, the relatively large day/night temperature contrast may be explained by recent results from \ct{Zhang2016}, which found that the day/night contrast generally increases with atmospheric metallicity in GCMs. The one key caveat here is that the high dayside temperature is difficult to explain if 55 Cancri e has a H$_{\mathrm{2}}$O of CO$_{\mathrm{2}}$ dominated atmosphere, since both these molecules have significant opacity in the IRAC 4.5 $\mathrm{\mu}$m bandpass, which means that these Spitzer observations should probe higher and colder parts of the atmosphere \cp{Demory2016b}. This caveat can be overcome though if there is a strong thermal inversion in the dayside atmosphere\cp{Demory2016b}. On the other hand, \ct{Tsiaras2016} recently announced the possible detection of atmospheric HCN on 55 Cancri e, which would also indicate the existence of volatile atmosphere on 55 Cancri e. 

So, based on the planet's mass and radius, we would probably conclude that 55 Cancri e currently has a significant water-dominated envelope, which has experienced a moderate amount of evaporation since it formed. However, there are other additional recent observations which complicate this picture and instead suggest that 55 Cancri e must have a relatively bare rocky surface. \ct{Demory2016a} appears to have detected large scale thermal variability in the dayside temperature between 2012 and 2013, possibly indicating the presence of the large scale surface volcanic activity. Moreover, \ct{RiddenHarper2016} recently announced the possible detection of a sodium and calcium exosphere escaping from 55 Cancri e, as might be expected for partially molten rocky planets with vigorous outgassing \cp{Schaefer2009,Valencia2010}. 

Taken as a whole, these new observations seem hard to reconcile, both with each other and with models of planetary structure. However, if we take them all at face value then they appear to suggest that we are seeing a relatively bare rocky planet, i.e., one without a deep volatile envelope, with an unusual equation of state and a relatively modest outgassed refractory atmosphere. One possibility suggested by \ct{Madhu2012}, is that if 55 Cancri e formed from material with a high C/O ratio it could possible have a lower density mantle composed of C or SiC rather than the normal silicates like Olivine and Perovskite found in the Earth's mantle. Such a carbon rich mantle would allow the planet to match the current mass and radius without needing a large volatile envelope. This proposal is somewhat supported by observations by \ct{Teske2013},which suggest a slightly enhanced stellar C/O of $0.78\pm0.08$, although \ct{Brewer2016} found a lower value of 0.53$\pm0.054$, consistent with Solar composition. At the moment though, any definitive conclusions about the composition of 55 Cancri e are premature and further work is needed from both observers and theorists to confirm these recent observations and to further and model characterize the planet's variability and possible exosphere.

\section{Further Questions and Future Observations}

These results make a clear prediction for the compositions of low-mass USP planets. These planets should be unable to retain low-metallicity H/He envelopes against photo-evaporation, and if these planets do, in general, form from water-poor material well within their stars' snow-lines then they should have bulk densities that are consistent with pure rocky compositions. Indeed, for the highly-irradiated transiting planets with well measured masses, this does seem to be the case \cp[e.g.,][]{Dressing2015, Buchave2016, Lopez-Morales2016}, with the possible exception of 55 Cancri e, as discussed above. However, due to the faintness of most {\it Kepler} target stars and the difficulty of obtaining precise mass measurements with radial velocity, the sample of USP planets is currently quite limited \cp{Dressing2015}. The observed lack of USP planets with radii $>$2 $R_{\mathrm{\oplus}}$ \cp{Sanchis-Ojeda2014,Lundkvist2016}, appears to rule out a large population of planets with substantial water envelopes that are a large fraction of their mass. However, given the lack of knowledge about their masses, it is possible that some planets with radii $\sim$1.5-2 $R_{\mathrm{\oplus}}$ may host modest water envelopes, comparable to the $\approx8\%$ envelope predicted to exist on 55 Cancri e. Consequently, it will be extremely important to acquire additional mass measurements for highly irradiated planets in this size range, in order to test the generality of this result. This should soon be possible since next year NASA is scheduled to launch the Transiting Exoplanet Survey Satellite (TESS), which will provide a much larger sample of super-Earth and sub-Neptune sized transiting planets around bright nearby stars that are easier to follow-up with RV \cp{Ricker2014}. Likewise, next year ESA and the Swiss Space Office plan to launch the CHEOPS mission \cp{Fortier2014}, which will look for transits from known RV planets and may also yield USP planets in the relevant size range.

Of course, if any transiting USP planets are confirmed with radii $>$2 $R_{\mathrm{\oplus}}$, these would be of extreme interest. Currently, no such planets are known, however many of the USP planet candidates have quite uncertain stellar parameters. Moreover, it is possible that current or upcoming transit surveys like K2 \cp{Howell2014}, TESS \cp{Ricker2014}, and CHEOPS \cp{Fortier2014} may yet uncover such planets, especially since these surveys will include more young stars where photo-evaporation may still be ongoing, unlike the old stellar population traced by {\it Kepler}. As a result, it is worth searching carefully through current and upcoming data for $\gtrsim$2 $R_{\mathrm{\oplus}}$ on orbits receiving $\gtrsim1000$ $F_{\mathrm{\oplus}}$, and if any such planets are found, they should be followed up intensely.

While this study sheds some valuable insight on the origin of the shortest-period low-mass planets found by {\it Kepler}, these planets are at the extreme end of the planet distribution and represent only a small fraction of the overall low-mass planet population \cp{Howard2012, Sanchis-Ojeda2014}. The vast majority of currently known sub-Neptune-sized planets are on more moderately irradiated orbits receiving $\sim10-1000$ $F_{\mathrm{\oplus}}$, where H/He envelopes can sometimes be retained against photo-evaporation and it is therefore more challenging to break the degeneracy between H/He and steam envelopes \cp{Rogers2010b}. In this region, more work is needed to characterize the possible compositions of these planets. One possibility is of course to directly constrain the abundance of these planetary atmospheres through atmospheric transmission spectroscopy and this is currently a focus of a great deal of observational work \cp[e.g.,][]{Berta2012, Kreidberg2014, Knutson2014, Fraine2014}. However, these observations are extremely challenging and their interpretation is often complicated by the possibility of high-altitude clouds \cp[e.g.,][]{Morley2013,Kreidberg2014}. Moreover, even is these planets are found to have low-metallicities in the observable portion of their atmospheres, this does nothing to rule out the possibility that they have water rich interiors or cores deeper down \cp{Fortney2013,Mordasini2013}. Therefore it is also important to search for potentially diagnostic features in the overall radius-flux distribution of planets. 

\section{Summary}

Using models of planetary thermal evolution and atmospheric photo-evaporation, I have examined the possible origin and evolution of low-mass ultra-short-period planets and compared these to current observations. Below I summarize the key results of this study.

\begin{itemize}

\item Low-mass planets with solar composition H/He envelopes receiving $>$1000$\times$ the irradiation of the Earth are incredibly vulnerable to atmospheric photo-evaporation, such that there should be essentially no non-giant planets with low-metallicity envelopes on USP orbits.

\item Consequently, photo-evaporation of such low-metallicity envelopes provides a natural explanation for the observed lack of sub-Neptune sized planets on USP orbits, here referred to as the "evaporation desert."

\item Conversely, however, very high metallicity or pure water envelopes should be much more resistant to erosion by photo-evaporation, and so if such planets existed, then a substantial fraction of them should retain their envelopes and be $>$2 $R_{\mathrm{\oplus}}$, even on USP orbits.

\item USP planets that are 2-4 $R_{\mathrm{\oplus}}$ are not seen in the {\it Kepler} sample, indicating that these planets formed from predominately water-poor material.

\item 55 Cancri e, is the one possible exception to this rule. However, current observations are difficult to interpret and further work is need to determine whether or not it is a bare rocky planet or a water-rich sub-Neptune.

\end{itemize}

These results are a key step in our quest to understand the origin of USP planets, and by extension the formation of low-mass planets in general. Moreover, this shows that models of photo-evaporation when combined with studies of the radius, mass, and orbital distribution of planets provide a valuable tool for constraining how planets form.

\section*{Acknowledgements}
I would like to thank the HARPS-N team, Jonathan Fortney, Tiffany Kataria, Laura Kreidberg, George McDonald, Ruth Murray-Clay, Christoph Mordasini, Roberto Sanchi-Ojeda, James Owen, Ken Rice, Johanna Teske, and Angie Wolfgang for their comments and input. This research has made use of the Exoplanet Orbit Database and the Exoplanet Data Explorer at exoplanets.org." The research leading to these results also received funding from the European Union Seventh Framework Programme (FP7/2007-2013) under grant agreement number 313014 (ETAEARTH).

\bibliographystyle{mnras}
\bibliography{USP_MNRAS_arxiv9.bib}

\bsp	
\label{lastpage}
\end{document}